\documentclass[pdflatex,sn-mathphys-ay]{sn-jnl}

\usepackage{graphicx}%
\usepackage{multirow}%
\usepackage{amsmath,amssymb,amsfonts}%
\usepackage{amsthm}%
\usepackage{mathrsfs}%
\usepackage[title]{appendix}%
\usepackage{xcolor}%
\usepackage{textcomp}%
\usepackage{manyfoot}%
\usepackage{booktabs}%
\usepackage{algorithm}%
\usepackage{algorithmicx}%
\usepackage{algpseudocode}%
\usepackage{listings}%

\theoremstyle{thmstyleone}%
\theoremstyle{thmstyletwo}%

\theoremstyle{thmstylethree}%

\begin{document}

\title[Near resonant 2D soliton interaction]{ {Experimental and numerical simulations of n}ear resonant 2D soliton interaction in shallow water}


\author[1]{\fnm{Thibault} \sur{Leduque}}\email{thibault.leduque@gmail.com}

\author[2]{\fnm{Fred} \sur{Couderc}}\email{couderc@math.univ-toulouse.fr}

\author[1]{\fnm{Maxime} \sur{Kaczmarek}}\email{maxime.kaczmarek@univ-grenoble-alpes.fr}

\author[1]{\fnm{Eric} \sur{Barth\'elemy}}\email{Eric.Barthelemy@grenoble-inp.fr}

\author[1]{\fnm{Nicolas} \sur{Mordant}}\email{nicolas.mordant@univ-grenoble-alpes.fr}

\author*[1]{\fnm{Herv\'e} \sur{Michallet}}\email{herve.michallet@univ-grenoble-alpes.fr}

\affil*[1]{\orgdiv{LEGI UMR 5519}, \orgname{Univ. Grenoble Alpes}, \orgaddress{\street{CNRS}, \city{Grenoble-INP}, \postcode{F-38000}, \state{Grenoble}, \country{France}}}

\affil[2]{\orgdiv{IMT UMR 5219}, \orgname{Univ. Toulouse}, \orgaddress{\street{CNRS}, \city{INSA}, \postcode{F-31077}, \state{Toulouse}, \country{France}}}


\abstract{
Experiments were carried out in   {a} large 
wave basin ($27\times 30$\,m$^2$) to study the interactions of non-linear waves in shallow water (0.35\,m deep). A stereoscopic technique was used to measure the waves accurately over a vast area ($180$\,m$^2$). The numerical code {Tolosa-lct}, designed for simulating wave propagation in shallow water, reproduces the observations and extends the analysis for larger domains.
The diffraction of oblique solitons is discussed. Mach interaction of two solitons occurs for some given incidence and amplitudes, producing a third soliton.
A focus is made on the near resonant interaction of a soliton with a vertical wall. An example of a regular Mach interaction is shown, where a steady solution is achieved experimentally, with a maximum amplitude amplification of 3.2 at the wall.
}




\maketitle

\section{Introduction}
The soliton is a mathematical object, localized in time and space, which preserves its identity after multiple interactions. 
The term `soliton' was coined by \citet{zabusky1965interaction} to define this remarkably stable object. 
It can be interpreted as a localized hump on the free surface in shallow water 
\citep[ {such an observation reported a long time ago by}][]{russell1845report} 
or as a modulation of the carrier wave of wave groups in intermediate water
\citep[][e.g.]{chabchoub2013experimental}
to deep water 
\citep[][e.g.]{fache2024interaction}
but also in other dispersive media such as in optic fibers 
\citep[][e.g.]{copie2025controlled}.
The soliton is a solution of  non-linear equations such as 
 {the Korteweg-de Vries (KdV) equation} \citep{Korteweg1895} or the nonlinear Schr\"odinger equation. 
Each soliton interaction causes a phase shift, but each soliton returns to its initial shape after interaction. This is rigorously true with KdV, when the two solitons propagate in the same direction (1D), at the surface of a fluid in shallow water (where the characteristic width of the soliton - its mass divided by its amplitude - is large in front of the water depth) for example. This makes KdV an integrable equation that conserves mass, momentum, energy and an infinity of other higher-order moments, so that the description of the system can be reduced to a few quantities (soliton amplitudes and phases) whereas it would require an infinite number of parameters in Fourier space \cite[e.g.]{Osborne,redor2019}. 

From an experimental point of view, in shallow water, it has been shown that the \cite{rayleigh1876} solution is well suited to produce solitons with little dispersive trailing waves with a piston-type wavemaker \citep{Guizien2}.
 {
The free surface displacement of a soliton is
\begin{equation}
\eta=a\, \text{sech}^2\left[\beta(x-ct)/2\right]\,,
\end{equation}
where $a$ is the amplitude of
the soliton, $c$ the phase speed, $\beta$ the shape factor, $x$ the space
coordinate and $t$ time. Here the soliton is a localized symmetric
hump traveling in one direction at constant speed. The parameters of 
the KdV solution fulfill:
\begin{equation}
c=\sqrt{gh_0}(\epsilon/2+1)
\end{equation}
and
\begin{equation}
\beta=\frac{\sqrt{3\epsilon}}{h_0}\,,
\end{equation}
while the
parameters of 
the Rayleigh solution fulfill:
\begin{equation}
c=\sqrt{gh_0(\epsilon+1)}
\end{equation}
and
\begin{equation}
\beta=\sqrt{\frac{3\epsilon}{{h_0^2}(\epsilon+1)}}\,,
\end{equation}
where ${h_0}$ is the quiescent water depth, $\epsilon$ the reduced amplitude $\epsilon=a/{h_0}$ and $g$ the gravity acceleration.
The $\beta$ parameter is a measure of the exponential
decay of the trailing edges of the soliton. 
The Rayleigh soliton becomes wider than the KdV soliton with increasing amplitude. Rayleigh profiles were shown to be close to experimental profiles measured in a wave flume by \citet{redor2020}, and corresponding celerities match well.
}
\cite{leduquePhD2024} also noted that the experiments of \cite{li2011mach} better fit Rayleigh than KdV soliton profiles.  

When solitons  propagate in  {different} directions (2D), interactions become much richer \cite[e.g.]{nicolas1993,kodama2010kp,wang2012oblique,choi2022numerical}. 
The  {Kadomtsev-Petviashvili (KP)} system of equations is designed to describe such situation for small angular spreading \cite[]{kadomtsev1970stability,kodama2016kp}.
The case of the interaction of an oblique soliton with a vertical wall can be seen as the interaction of two oblique indentical solitons. 
\citet{miles1977resonantly}  defined this interaction as  `Mach interaction', where a parameter depending on incidence angle and amplitude determines if the interaction is regular or irregular. 
In the first case, the hump or `stem' at the wall is steady. 
In the irregular interaction case however, the stem width increases continuously. 
At the transition between the two cases a resonance is achieved where the stem amplitude can reach four times the amplitude of the incident soliton, a result confirmed numerically by \citet{knowles2019fourfold}. 

The aim of this article is to discuss ways of reproducing relatively simple interactions between two or three solitary waves in two horizontal dimensions (2D). Experiments in a wave tank have been carried out, pointing out the inherent limitations linked to the constrained space domain and the dissipation (section \ref{exp}). Numerical simulations (section \ref{simu}) reproduce the experiments very well (section \ref{valid}). The Mach interaction of a soliton against a wall is analyzed in more detail in section \ref{intMach}. 

\section{Methodology} 
\subsection{Experiments}\label{exp}

\begin{figure}
\centering
{{\includegraphics[width=85mm]{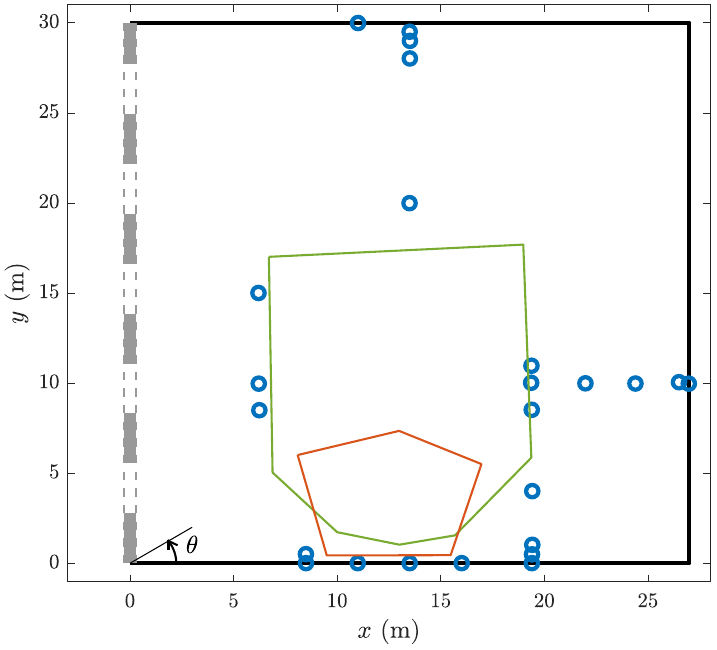}} }
\caption{ {Schematic top view of the wave tank: wavemakers on the left 
(they can move $\pm 30$\,cm with respect to their mean position $x=0$)},
areas covered by the stereoscopic systems (pair of {\sc pco} cameras in red and pair of {\sc jai} cameras in green), positions of wave gauges (blue circles).
\label{bassin_LHF24_tibo}
}
\end{figure}

Experiments were carried out in the large wave tank at the  {Laboratoire Hydraulique de France (LHF, Artelia company, located in} Pont-de-Claix, France). The 30\,m$\times$27\,m basin is equipped with a wave-maker consisting of 60 movable vertical paddles 50\,cm wide which can be controlled independently. 
Their maximum stroke is 60\,cm.
The other three sides of the basin are rigid, wave-reflecting vertical walls (see Fig.\,\ref{bassin_LHF24_tibo}). The bottom of the basin is horizontal on average, but irregularities of up to $\pm 5$\,mm can occur in places. The water depth at rest is ${h_0}=0.35$\,m for all experiments.

The wavemaker can generate solitary waves very close to the theoretically predicted  {Rayleigh} solitons. This was verified using a stereo-video technique by 
 \cite{leduque2024}. Capacitance-type wave gauges were also used to measure free surface displacements at various points in the tank. 

The area covered by the cameras, about 180\,m$^2$, is shown in Fig.\,\ref{bassin_LHF24_tibo}. By carrying out two symmetrical experiments with identical forcing, 
by changing prescribed motion of paddle $n$ ($n=[1,2,...60]$) to paddle $60-n+1$,
and combining their results, we claim to cover around 320\,m$^2$ of the basin. Two sets of two cameras were deployed, rated {\sc pco} (5.5 Mpixels, 16 bits) and {\sc jai} (12 Mpixels, 8 bits) in Fig.\,\ref{bassin_LHF24_tibo}. They record up to 20 images per second. Stereoscopic reconstruction provides free-surface elevations 
with a resolution of approximately 5 cm on the horizontal axes, $x$ and $y$, and a few millimeters on the vertical axis $z$ \citep{leduque2024}. The cameras cannot measure within 0.5\,m of the wall $y=0$. Wave gauges were fixed at 0.01\,m from this wall (at $x=8.5$, 11, $13.5$, $16$, $19.4$\,m) to capture wave amplification against it. A detailed description of the experimental set-up is presented in \cite{leduquePhD2024}.

\begin{figure}
{{\includegraphics[width=127mm,viewport=94 252 462 595,clip]{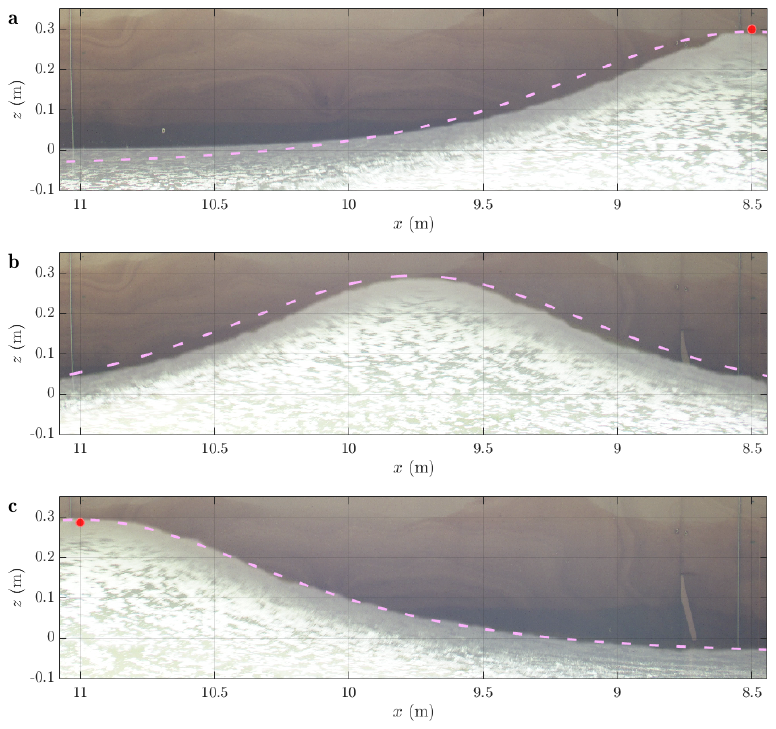}} }
\caption{Video frames (from a camera located at $x=10$\,m, $y=30$\,m) of the passage of a wave against the wall $y=0$, 
resulting from the generation of an oblique soliton 
($\epsilon=0.34$, $\theta=-38^\circ$). 
The wave propagates from right to left in this representation.
Time interval between two successive images: $\Delta t=0.43$\,s. 
Measurements of the maximum free surface elevation with wave gauges fixed to the wall (red dots).
The dashed line is Miles strong interaction \cite[eq. 5.11]{miles1977obliquely}.
\label{cambobo}}
\end{figure}

The free surface is covered with floating particles that form patterns  required to enable stereoscopic reconstruction \citep{leduque2024,leduquePhD2024}. 
In Fig.\,\ref{cambobo} are shown video frames of the passage of a wave at the wall $y=0$ recorded with another video-camera. 
This was meant to validate the  calibration of the wave gauges.
The maximum wave amplitude (red dots in Fig.\,\ref{cambobo}{\bf a} and {\bf c}) is estimated within a few millimeters for this 30\,cm wave height.
This also leads to further observations. 
The wave profile on the wall is rather close to
Miles solution for strong symmetric interactions  
\cite[eq. 5.11]{miles1977obliquely}, 
plotted in dashed.
 {This solution will be written explicitly in section \ref{intMach} along with details on Mach interaction theory.}

Of note, the theoretical profile has been shifted of -4\,cm for  matching the measured stem amplitude, which is nearly constant over this 2.5\,m travel.
Miles' strong interaction solution matches very well the rear face of the wave (Fig.\,\ref{cambobo}{\bf c}). 
Instabilities in the front of the wave can be seen in Fig.\,\ref{cambobo}{\bf a}. These can be attributed to the presence of the wave gauge.
Miles' solution does not match very well the front face of the wave
(Fig.\,\ref{cambobo}{\bf a}), nor the wave profile 0.43\,s later 
(Fig.\,\ref{cambobo}{\bf b}). This illustrates that this stem is not fully steady with this type of soliton generation.
This aspect will be discussed further in section \ref{intMach}.

\subsection{Numerical simulations}\label{simu}

Along with these experiments, numerical simulations of soliton interactions were carried out using the {Tolosa-lct} code part of the Tolosa suite
({tolosa-project.com}). 
The Tolosa code suite was originally developed to predict coastal submersion over complex topographies with low computational cost using unstructured meshes. Its new version, Tolosa-lct, extends this framework to  {fully nonlinear and dispersive wave propagation}.

{ 
The underlying model is not the Serre--Green--Naghdi system itself, but the hyperbolized Green--Naghdi model derived by \cite{richard_Tolosa} from the Euler equations integrated over the vertical under a weak-compressibility assumption. Three variables are added to the classical Saint-Venant system: the depth-averaged vertical velocity $w$, the non-hydrostatic pressure $p_a$, and a strain-like vector ${\bf s}\sim\nabla h$ which, following the approach of \citet{bonneton2011splitting}, improves the dispersive properties through a coefficient $\alpha_d$. Over the flat and fixed bottom of the basin, and in the absence of breaking, the system reads
\begin{equation}
\left\{
\begin{aligned}
&\partial_t h + \nabla\cdot(h{\bf u}) = 0\,,\\[6pt]
&\partial_t (h{\bf u}) + \nabla\cdot(h{\bf u}\otimes{\bf u}) = -\,g h \nabla h - \nabla(h\,p_a) + {\bf f}\,,\\[6pt]
&\partial_t (h w) + \nabla\cdot(h w\,{\bf u}) = \dfrac{3}{2}\,p_a + \dfrac{\alpha_d-1}{2\alpha_d}\,g h^2\,\nabla\cdot{\bf s} + 4\,\dfrac{\alpha_d-1}{\alpha_d^2}\,w^2 - (\alpha_d-1)\,h^2 \det(\nabla{\bf u})\,,\\[6pt]
&\partial_t (h p_a) + \nabla\cdot(h p_a\,{\bf u}) = -\,c_a^2\left(2 w + \alpha_d\,h\,\nabla\cdot{\bf u}\right)\,,\\[6pt]
&\partial_t (h{\bf s}) + \nabla\cdot(h{\bf s}\otimes{\bf u}) = 2 h \nabla w + \dfrac{2}{\alpha_d}\,w\,{\bf s} - {\bf s}\cdot\left(h\nabla{\bf u} - h\,(\nabla\cdot{\bf u})\,{\bf I}\right)\,,
\end{aligned}
\right.
\end{equation}
where $h$ is the total water depth, ${\bf u}$ the depth-averaged horizontal velocity, ${\bf f}$ the bottom friction force (section\,\ref{valid}) and $c_a$ a finite artificial sound speed. The dispersive effects are carried by the acoustic $(w,p_a)$ sub-system, whose waves travel at the speed $c_a\gg\sqrt{g h_0}$, $h_0$ being the quiescent water depth. A splitting strategy decouples these ``fast" acoustic waves from the ``slow" gravity waves, the acoustic sub-system being sub-iterated within each Saint-Venant time step. The artificial Mach number $M_a=\sqrt{g h_0}/c_a$ is set to $0.2$ here, i.e. $c_a\simeq 9.3$\,m\,s$^{-1}$. In the limit $c_a\to\infty$, the linear dispersion relation over a flat bottom reads
\begin{equation}
\omega^2 = g h_0 k^2\,\dfrac{1+(\alpha_d-1)\,(k h_0)^2/3}{1+\alpha_d\,(k h_0)^2/3}\,.
\end{equation}
For $\alpha_d=1$ it reduces to the Serre--Green--Naghdi relation, which matches the Airy relation $\omega^2=g k\tanh(k h_0)$ at first order in $(k h_0)^2$ only, whereas the value $\alpha_d=1.2$ used here gives the Pad\'e (2,2) approximant of the Airy relation, accurate up to $O\left((k h_0)^4\right)$.

The interest of this formulation is that the Serre--Green--Naghdi equations require, at each time step, the inversion of an elliptic operator over the whole domain, which is costly and scales poorly in parallel, whereas the present system is purely hyperbolic, hence explicit and local. It is discretized by a finite-volume scheme on unstructured meshes with message-passing interface (MPI) parallelism, the extra cost being limited to a factor of 4 to 6 with respect to a standard Saint-Venant solver. The model is fully nonlinear, with no restriction on the relative amplitude, but it remains depth-averaged: it assumes a single-valued free surface and an irrotational flow of an inviscid fluid, and therefore describes neither overturning waves nor the wave boundary layer. Wet--dry fronts are handled by the code, and the unresolved dissipative processes are modeled by the bottom friction law of section\,\ref{valid}.
}

For the present study, the code has further been equipped with an  {Arbitrary Lagrangian-Eulerian (ALE)} formulation that reproduces the prescribed motion of the wavemaker, allowing a direct comparison with the experiments in terms of soliton generation, soliton--soliton interactions, and reflection against the basin walls.
 {
Following a mesh convergence study, we set the resolution to 2\,cm, which corresponds to a mesh of $1350\times1500$ cells. The ALE formulation requires the use of a deformable mesh, which we chose to apply to the first 10 meters. We verified that this did not affect the accuracy of the simulation.
}

\subsection{Validation: interaction of two solitons }
\label{valid}

\begin{figure}
{{\includegraphics[width=127mm,viewport=102 290 538 562,clip]{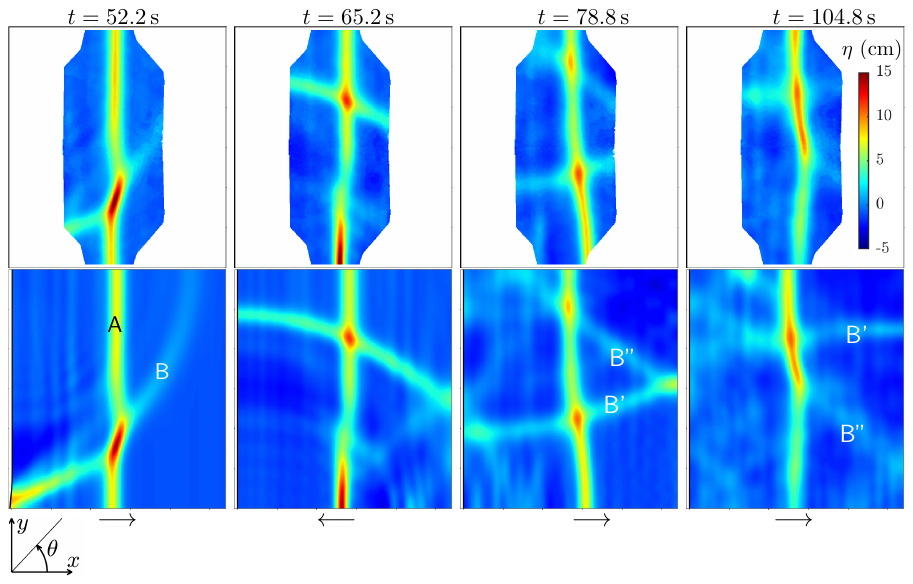}} }
\caption{
Stereoscopic measurements (top row) and simulation {Tolosa-lct} (bottom row) of the free-surface elevation at different stages following the generation of two solitary waves. {\sf A}: amplitude relative to water depth $\epsilon=0.3$, incidence $\theta=0$ generated at $t=19.7$\,s and {\sf B}: $\epsilon=0.13$, $\theta=-60^\circ$ at $t=39.5$\,s  
($t=0$ is the start of the backward motion of the paddles, indicated times are the start of the forward push).
{\sf B'} and {\sf B''} are the main resultants of {\sf B} interactions. 
Each frame covers 27\,m in $x$ and 30\,m in $y$. 
The arrows indicate the direction of propagation of the main wave resulting from the soliton {\sf A}.
\label{compa_2514}
}
\end{figure}

These measurement and calculation tools need to be compared. 
An example is shown in Fig.\,\ref{compa_2514} for the generation of two solitary waves.
A first wave ({\sf A}) is generated at $t=19.7$\,s with all wavemaker paddles in phase (incidence $\theta=0$) and reflects against the wall at $x=27$\,m then against the wavemaker  at $x=0$, while a second wave ({\sf B}) is generated with a set incidence of $-60^\circ$ at $t=39.5$\,s. 
With angles, the aperture of the wavemaker restricts the length of the line soliton. This induces diffraction, the soliton bends and its amplitude decays along the line crest. This effect is more pronounced at large angles. 
Wave {\sf B}
interacts with the first soliton {\sf A} (Fig.\,\ref{compa_2514} at $t=52.2$\,s) and reflects against the wall at $y=0$. A second interaction is shown at $t=65.2$\,s: the main wave resulting from {\sf A} is not uniform along $y$, showing a maximum near the wall $y=0$ while the wave resulting from {\sf B} is strongly curved. The main wave of incidence $\theta\approx 0$ reflects at the wavemaker , then interacts with two solitary waves of incidence $\theta\simeq -80^\circ$ ({\sf B'}) and $\theta\simeq -130^\circ$ at $t=78.8$\,s ({\sf B"}). These were created by the interaction of the {\sf B} wave with the walls at $y=30$\,m and $x=27$\,m. At $t=104.8$\,s, the main resultants of these two waves {\sf B'} and {\sf B''} can be seen interacting again with the $\theta\simeq 0$ wave from {\sf A}. 

Agreement between experiment and simulation is remarkably good after several interactions of these solitary waves. The simulation over the whole basin (bottom row Fig.\,\ref{compa_2514}) clearly illustrates the richness of the waves' interactions with each other and with the walls. 

A soliton interacting frontally with a wall dissipates its energy by producing dispersive waves 
 {\cite[e.g.]{Renouard1985}. This type of interaction is similar to the 1-D head-on collision of two solitons which is not fully elastic, also known as weak interaction in contrast to the strong interaction of two solitons propagating in the same direction \citep[see][e.g.]{redor2020}.
These dispersive waves are the vertical stripes
seen on the left}
of Fig.\,\ref{compa_2514} at $t=52.2$\,s. 
 {
In contrast, the interaction between the waves within the basin at later times look more elastic.
}
At $t=104.8$\,s, wave crests are clearly visible, which could be related to solitons, with a tendency to regularize along $x$, $y$, or in the diagonal of the basin.

The solitons, once generated at the wavemaker, interact with the walls and dissipate through bottom friction. One of the difficulties in simulating experiments is to model this bottom friction correctly. 
 {
{Like the Serre--Green--Naghdi equations it approximates, the model solved by Tolosa-lct assumes an irrotational flow of an inviscid fluid.} The dissipation by bottom friction is therefore parametrized. The resulting force on the flow consists of two terms that simulate the dissipation in both laminar and turbulent boundary layers, as:
\begin{equation}
{\bf f} = -C_l {\bf u} - C_q \|{\bf u}\| {\bf u} \,.
\end{equation}
The model was evaluated against the measured decay of a frontal soliton traveling back and forth in the basin by \cite{leduquePhD2024}.
}
 {
In the present computation, we take $C_l=2\,10^{-3}$ m\,s$^{-1}$ and $C_q=0$ as a good compromise to model the dissipation of both small- and large-amplitude waves \cite[see details in][sections 2.6.2 and 3.1.1]{leduquePhD2024}. 
}

\begin{figure}
{{\includegraphics[width=127mm,viewport=100 295 522 561,clip]{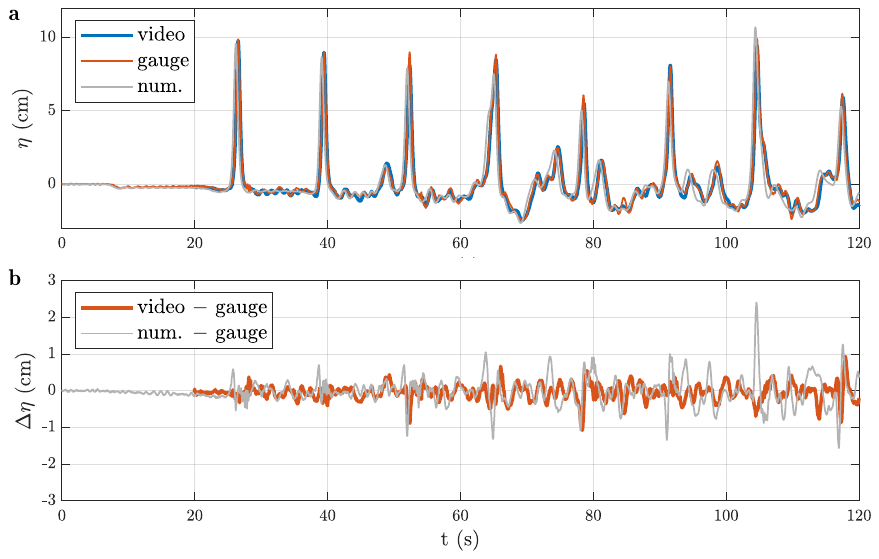}} }
\caption{
{\bf a}
Free surface elevations following the generation of two solitons for two symmetrical experiments ({\sf A}: $\epsilon=0.3$, $\theta=0$, $t=19.7$\,s; {\sf B}: $\epsilon=0.25$, $\theta=\pm 60^\circ$, $t=39.5$\,s), measured by a capacitance gauge located at ($x=13.5$\,m, $y=20$\,m), by video at the symmetrical point ($x=13.5$\,m, $y=10$\,m) and numerical simulation {Tolosa-lct}. {\bf b} Differences 
between gauge and video measurements
(in red) and between video measurement and simulation (in grey).
\label{C_V_T_2514}
}
\end{figure}

 {The symmetric case was performed,} 
so that following the 
first frontal soliton, an oblique soliton of incidence  $\theta = + 60^\circ$ was generated $19.8$\,s later. This procedure was used to compare the water elevation measurements of the stereo video system with a capacitance gauge located in the part of the basin out of the field of view of the cameras.

In Fig.\,\ref{C_V_T_2514}{\bf a} are compared the time series measured with the wave gauge and the symmetrical point obtained from the stereo video, for the two symmetrical experiments.
This comparison verifies that the stereoscopic measurements are very close to those of the gauge, and that our experiments are reproducible by symmetry. 
The differences are plotted in Fig.\,\ref{C_V_T_2514}{\bf b}, confirming that the agreement between video and gauge is very good. Despite refraction effects induced by bottom irregularities and other generation defaults, the agreement remains within a few millimeters. Discrepancies result from slight wave phase differences. 

The signal extracted from the numerical simulation {Tolosa-lct} is also plotted. 
{Tolosa-lct} reproduces the global dynamics of the interactions well, the differences with the gauge at the same point are mainly induced by a slight phase delay at each soliton passage.
This small discrepancy 
might result from a combination of several sources of error, such as bottom irregularities and limited accuracy of the mean level estimation, or the flawed dissipation model in the numerical simulation. Another source of discrepancy is that there is no watertight seal between the paddles of the wavemaker: there is a weak flow in between especially during the receding phase of the motion of the paddles. In contrast, the wavemaker in the simulation is a deformable rigid wall. 

To characterize the dynamics of 
highly non-linear wave interactions, with each other and with vertical walls, 
and in particular to reproduce the dispersive waves generated, the {Tolosa-lct} code seems nevertheless well suited.

\section{Mach interaction} 
\label{intMach}

\subsection{Theory}

The maximum amplitude reached in the resonant interaction of a soliton with a wall has long been a subject of controversy \citep{perroud1957solitary,miles1977resonantly,li2011mach,kodama2016kp,knowles2019fourfold}. \citet{miles1977resonantly} was the first to theorize this interaction, which predicts a maximum amplitude at the wall of 4 times the amplitude of the incident soliton, provided that reduced amplitude ($\epsilon_i=a_i/{h_0}$) and incident angle ($\theta_i$) of the soliton 
 {
are both small and}
satisfy 
\begin{equation}
\theta_i=\sqrt{3\epsilon_i}\,.
\end{equation}
The parameter 
\begin{equation}
\kappa_{m} = \frac{\theta_i}{\sqrt{3\epsilon_i}}
\label{kappaM}
\end{equation}
is introduced.
The steady solution of \cite{miles1977resonantly}, called `Mach interaction', predicts the resonance at $\kappa_{m}= 1$. 
A kinematic equilibrium is reached, where incident soliton, reflected soliton and stem all propagate at the same speed along the direction propagation of the interaction point. 
In the irregular interaction case ($\kappa_{m}<1$), also known as `Mach reflexion', the third soliton against the wall (or `stem') 
infinitely grows in width, along the steady solution propagating with a given angle to the wall.
In a regular interaction (for $\kappa_{m}>1$), the stem, which is not a soliton, has a constant width throughout propagation.
The amplification ($\alpha_w$) of the stem amplitude ($a_w$) relative to that of the incident soliton ($a_i$) is:
\begin{equation}
\alpha_w=\frac{a_w}{a_i}  =
    \begin{cases}
(1+\kappa_{m})^2 & \text{for $\kappa_{m}<1$}\\
\frac{4}{1+\sqrt{1-\kappa_{m}^{-2}}} & \text{for $\kappa_{m}>1$}\,.
    \end{cases}       
		\label{alpha_miles}
\end{equation}

\cite{kodama2016kp} have revisited the analysis of \cite{miles1977resonantly} 
and found that this maximum amplitude must be interpreted within the framework of a second-order KP theory. KP is a weakly nonlinear, weakly dispersive equation with little directionality ($\theta_i$ small), established by \cite{kadomtsev1970stability}. This second-order analysis, allowing larger angles,  gives a creditable account of experiments by \cite{li2011mach} and numerical simulations by \cite{funakoshi1980reflection} and \cite{tanaka1993mach}, extended to simulations by \cite{knowles2019fourfold}. In this analysis, the amplitude amplification of the soliton incident on the wall follows Miles' law as long as a virtual amplification $\hat\alpha$ predicted by KP at second order is applied.  
In this approach proposed by \cite{kodama2016kp} \cite[see][(1.3)]{knowles2019fourfold}:
\begin{equation}
\kappa= \frac{\sqrt{1+\sqrt{1+5\epsilon_i}}\tan\theta_i}{\sqrt{6\epsilon_i}\cos\theta_i}
\label{kappaK}
\end{equation}
and 
\begin{equation}
\hat\alpha = 
    \begin{cases}
\frac{\alpha_w(1+\sqrt{1+5\epsilon_i})}{(1+\sqrt{1+5\alpha_w\epsilon_i})\cos^2(\theta_i)} & \text{for $\kappa<1$}\\
\alpha_w & \text{for $\kappa>1$}
    \end{cases}        \,.
\label{alfachapo}
\end{equation}
For $\theta_i\ll 1$ and $\epsilon_i\ll 1$, $\kappa\sim \kappa_m$ 
 {
and $\hat\alpha\sim \alpha_w$.
}
 The resonance 
 {
theoretically}
 occurs  when $\kappa=1$
where $\hat\alpha$ can reach 4.
 {
The 
physical amplification 
$\alpha_w$
of the wave may remain lower than $\hat\alpha$ for irregular interactions ($\kappa<1$). 
}
Note that the correction proposed by \cite{kodama2016kp} (\ref{alfachapo}) presents a discontinuity for $\kappa=1$. 
The experimental observation doesn't seem to show any drastic discontinuity between regular reflection and the production of a third wave, as will be shown 
 {
in the next section. 
}

 {
In this framework, 
Miles solution of the free surface elevation for strong symmetric interactions  
\cite[eq. 5.11]{miles1977obliquely} reads:
\begin{equation}
\eta=4 \epsilon_i {h_0} \frac{1+e^{\delta}\cosh 2(x\cos\theta_i+{\textstyle\frac{1}{2}}\delta) \cosh 2y\sin\theta_i}
    {(\cosh 2(x\cos\theta_i+{\textstyle\frac{1}{2}}\delta) + e^{\delta} \cosh 2y\sin\theta_i)^2}  
		\label{milesstrong}
\end{equation}
with \cite[][eq. 5.6]{miles1977obliquely}:
\begin{equation}
\delta={\textstyle\frac{1}{2}}\log|1-\kappa^{-2}| 
		\label{delta}
\end{equation}
with our notations and taking $\kappa$ from (\ref{kappaK}). 
}

\subsection{Experiments}

\subsubsection{Single oblique soliton}

Experiments involving the interaction of a soliton with a vertical wall are generally carried out in a tank, in which a wall is installed at the desired angle to the wavemaker, thus defining a triangular domain. The generated soliton propagates in the direction perpendicular to the wave-maker and interacts with the wall. Our configuration is different, since the soliton is generated in a rectangular domain with an incidence $\theta_i$ that can be different from 0 by shifting the phase of each paddle appropriately \citep{leduquePhD2024,leduque2024}. 
Due to the wavemaker segmentation (50\,cm-wide paddles moving in the $x$-direction) and diffraction (see section \ref{valid}), 
there is some loss in the process so that, for large angles, 
the amplitude of the wave incident onto the wall is smaller than the amplitude of the target soliton. 
Here the reference amplitude is measured by a wave gauge at a location ($x=6.25$\,m, $y=8.50$\,m) which is not too far from  the wall $y=0$ and not too close  to the wave-makers (also at the boundary of the area covered by the cameras).

\begin{figure}
\centering
{{\includegraphics[width=107mm,viewport=100 280 513 563,clip]{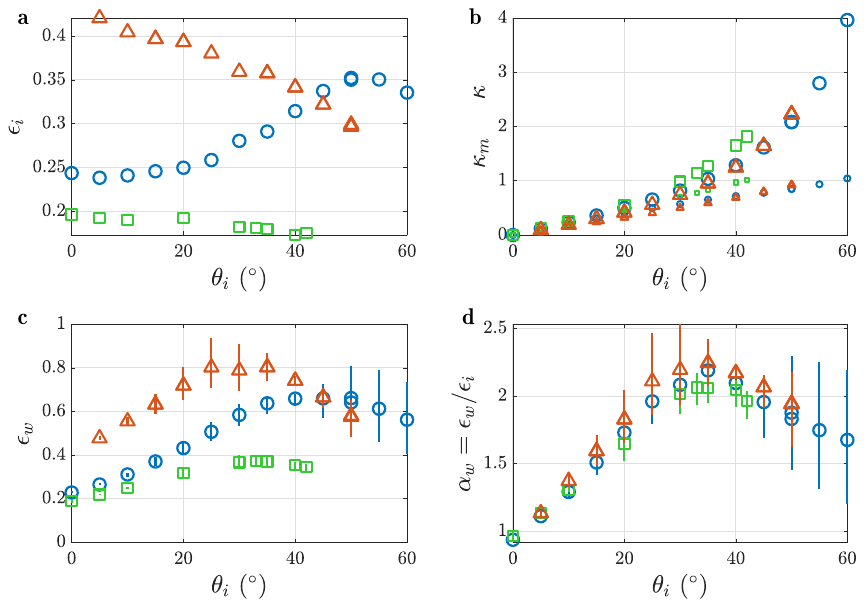}} }
\caption{
Measurements in Mach interaction for various incidence angles $\theta_i$ and three different series of runs (triangles, squares and circles): 
{\bf a} relative amplitude of incident soliton $\epsilon_i$ measured by the gauge located at ($x=6.25$\,m, $y=8.50$\,m);
{\bf b} Miles parameter (\ref{kappaM}) ($\kappa_m$: small symbols) and Kodama \& Yeh parameter (\ref{kappaK}) ($\kappa$: large symbols);
{\bf c} relative amplitude at the wall $\epsilon_w$ measured by four gauges located at $y=0.01$\,m and $x=8.50,\, 11.00,\, 13.50,\, 16.01$\,m; 
{\bf d} amplitude amplification at the wall $\epsilon_w/\epsilon_i$. 
In {\bf c} and {\bf d}
the symbols indicate the mean value, vertical bars indicate the minimum to maximum values of the four gauges.
\label{epsilon_kappa_aw_theta}
}
\end{figure}

Three sets of runs were conducted, by varying mainly the incidence angle of the soliton. In two cases (triangles and squares in Fig.\,\ref{epsilon_kappa_aw_theta}), 
 {
the target amplitude was fixed ($\epsilon_0=0.4$ and $\epsilon_0=0.25$, respectively). Due to the segmented wave-maker, there is some loss in the generation process and 
the measured reduced amplitude $\epsilon_i$ of the soliton entering the video-surveyed zone decreases with increasing $\theta_i$. 
The target amplitude $\epsilon_0$ is increased with increasing incidence}
in another series (circles), in order to better cover the parameter space ($\epsilon_i$, $\theta_i$).
Fig.\,\ref{epsilon_kappa_aw_theta} depicts the different incidence amplitudes (Fig.\,\ref{epsilon_kappa_aw_theta}{\bf a}), 
related Mach interaction parameters (Fig.\,\ref{epsilon_kappa_aw_theta}{\bf b}), maximum wave amplitude at the wall (Fig.\,\ref{epsilon_kappa_aw_theta}{\bf c}). 
As a result, with this set of parameters and generation procedures, the amplitude amplification at the wall depends mainly on the soliton incidence angle (Fig.\,\ref{epsilon_kappa_aw_theta}{\bf d}).

\begin{figure}
\centering
{{\includegraphics[width=107mm,viewport=107 350 524 503,clip]{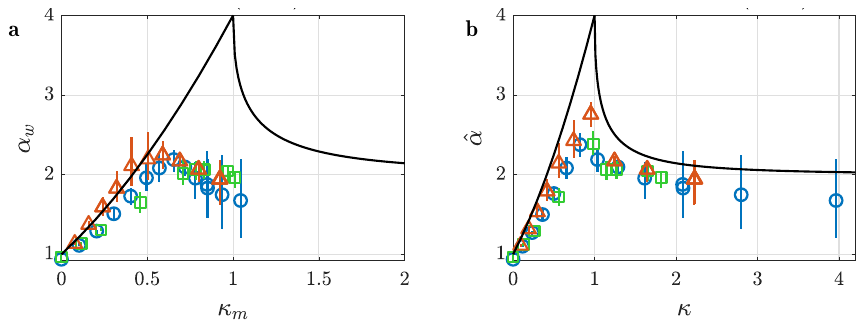}} }
\caption{
 {
Experimental results of {\bf a} the amplification factor $\alpha_w$ 
versus $\kappa_m$, the Miles parameter (\ref{kappaM}) 
and {\bf b} the KP amplification factor $\hat\alpha$ versus
\cite{kodama2016kp}'s higher-order KP parameter $\kappa$ (\ref{kappaK}). 
}
Generation of a single oblique soliton with various incidence angles and relative amplitudes measured at ($x=6.25$\,m, $y=8.50$\,m), same symbols as in Fig.\,\ref{epsilon_kappa_aw_theta}.
The symbols indicate the mean value, vertical bars indicate the minimum to maximum values measured with the four gauges located close to the wall ($y=0.01$\,m) for each run.
 {
In solid line is Miles' solution (\ref{alpha_miles}) in {\bf a}, the same formulation as in (\ref{alpha_miles}) replacing $\kappa_m$ by $\kappa$ and $\alpha$ by $\hat\alpha$ in {\bf b}.
}
\label{alfamanip}
}
\end{figure}

Fig.\,\ref{alfamanip} plots the amplification at the wall 
for these gauges measurements and compared to \cite{miles1977resonantly}  and \cite{kodama2016kp} theories. 
The main difference in the two approaches is the computation of the parameter that characterizes the type of interaction: irregular (with amplification of the stem width throughout the wall interaction) or regular. The threshold is $\kappa_m=1$ (\ref{kappaM}) for \cite{miles1977resonantly}  and modified to $\kappa=1$ (\ref{kappaK}) by \cite{kodama2016kp}. 
The second difference made by \cite{kodama2016kp} 
was to consider that the amplitude amplification for irregular interaction ($\kappa<1$) has to be considered in a second order KP framework, so that the amplification factor has to be corrected (\ref{alfachapo}).

Our measurements (Fig.\,\ref{alfamanip}) clearly show a decay of the amplitude amplification at the wall around $\kappa=1$.
The correction of the amplification factor might even improve the trend (Fig.\,\ref{alfamanip}{\bf b}). 
The measurements nevertheless do not match the theory, lying below the predictions. This might be attributed 
mainly to the distance needed to establish a steady solution.  

Despite these limitations, our experimental configuration can produce 
 {
solitons in quasi-steady interaction,
}%
as shown in the next section.

\begin{figure}
{{\includegraphics[width=127mm,viewport=97 373 536 483,clip]{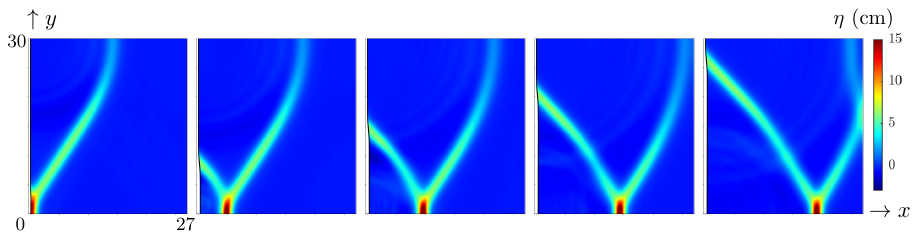}} }
\caption{
Simulation with {Tolosa-lct} of three-wave interaction {3S}: free surface elevation from stem generation, then one frame every 2\,s. Each frame covers 27\,m in $x$ and 30\,m in $y$. 
\label{Tolosa2244}
}
\end{figure}

\subsubsection{Three-solitons solution (3S)}

In an attempt to produce near resonant interaction established directly at the wavemaker, three solitary waves were generated in succession. A first soliton of incidence $\theta_i=-34^\circ$ is immediately followed by the production of a frontal soliton ($\theta_w=0^\circ$) on the first four wavemaker paddles ($y\leq 2$\,m). A third soliton is generated ($\theta_r=44^\circ$) to account for the wave reflected by the wall. 
This test is referred as {3S} in this paper. 
The numerical simulation of this test is shown in Fig.\,\ref{Tolosa2244}. The third generated wave limits the diffraction of the reflected wave produced by the wall interaction, but is not in equilibrium. The reflected wave tends to take an incidence $\theta_r'\simeq 30^\circ$, so that the interaction at the wall resembles a regular interaction.

\begin{figure}
{{\includegraphics[width=127mm,viewport=110 225 520 625,clip]{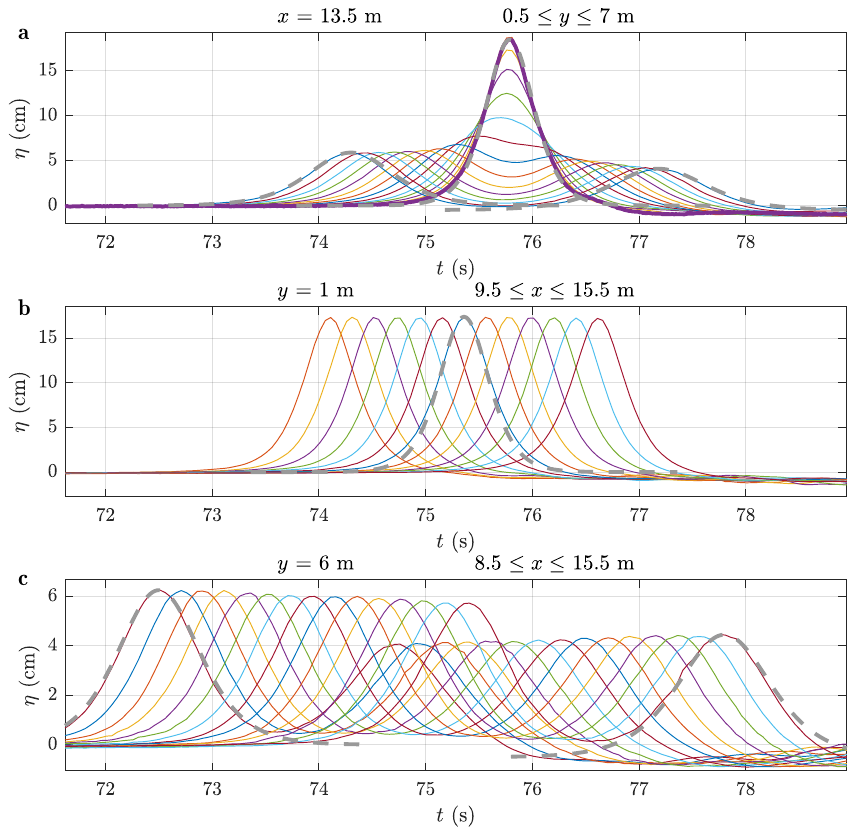}} }
\caption{
Stereoscopic measurements of the free surface elevation in  the three-wave wall interaction {3S} (thin lines). 
{\bf a} Time series every 0.5\,m along $0.5\leq y\leq 7$\,m at $x=13.5$\,m, {\bf b} every 0.5\,m along $9.5\leq x\leq 15.5$\,m at $y=1$\,m and {\bf c} along $8.5\leq x\leq 15.5$\,m at $y=6$\,m. 
The thick dashed lines are Rayleigh soliton profiles.
\label{MachInt_rayleigh2244}
}
\end{figure}

\begin{figure}
{{\includegraphics[width=127mm,viewport=0 0 570 245,clip]{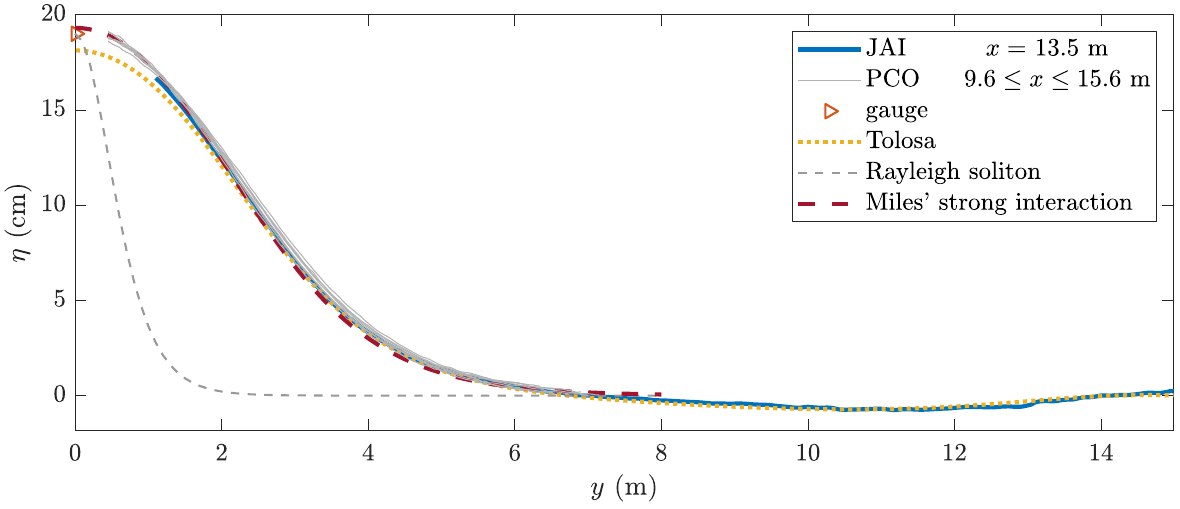}} }
\caption{
Profiles of the stem along $y$, experiment {3S}.
Stereoscopic measurements:
{\sc jai} profile at $x=13.5$\,m (thick solid line),
{\sc pco} profiles every 0.5 m in $x$ (thin solid lines).
The symbol ($\triangleright$) corresponds to wave gauge measurements at ($x=13.5$\,m, $y=0.01$\,m). 
Tolosa simulation (thick dotted).
The dashed grey line is a Rayleigh soliton profile that would propagate along $y$ (for reference). 
The thick dashed line is Miles solution  {(\ref{milesstrong}-\ref{delta})} for strong interaction for $\kappa=1.03$ ($\epsilon_i=0.171$ and $\theta_i=-30.4^\circ$).
\label{MachInt_stem}
}
\end{figure}

The stem produced in this interaction is remarkably stable, as shown by the time series extracted from the video measurements plotted in Fig.\,\ref{MachInt_rayleigh2244}. The $\eta(t)$ profiles of the incident wave, the stem and the reflected wave are very close to Rayleigh soliton profiles. 
 {
This means that these three waves all travel together as individual solitons along their own direction of propagation.
}%
At 6\,m from the wall, the incident wave tends to slightly decay in amplitude, while the reflected wave amplitude slightly increases (Fig.\,\ref{MachInt_rayleigh2244}{\bf c}). 
This observation is therefore qualified as quasi-steady.

The transverse profile (along $y$) of the stem varies very little throughout the area covered by the cameras, see Fig.\,\ref{MachInt_stem}. As expected, the stem is much longer than a characteristic soliton width, 
 {
somehow 
matching the strong interaction profile deduced 
from \cite{miles1977obliquely} for which $\kappa_m$ is replaced by $\kappa$  (\ref{milesstrong}-\ref{delta}). 
}

\begin{figure}
\centering
{{\includegraphics[width=80mm,viewport=110 270 485 585,clip]{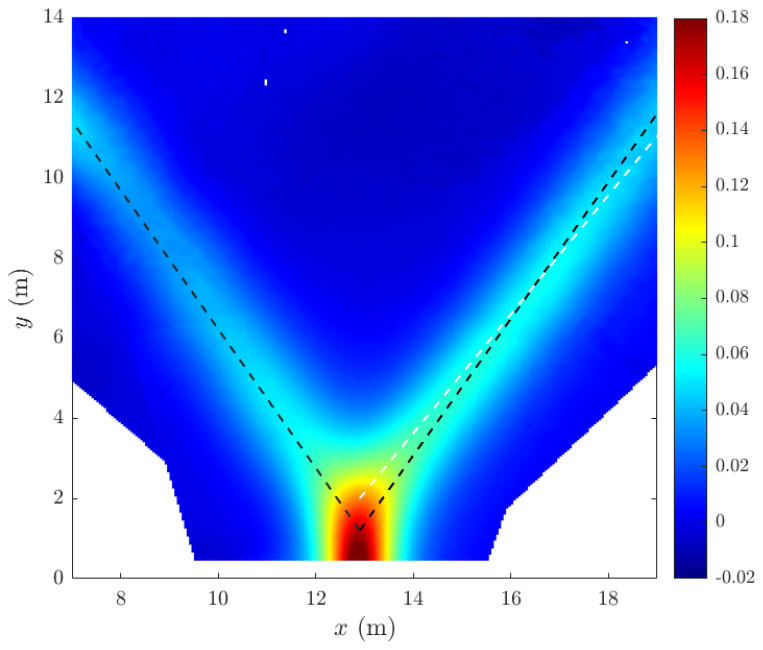}} }
\caption{
Snapshot of the free surface displacement (m) in the three-wave interaction {3S} measured by stereoscopy. Prescribed angle of the first generated oblique soliton (white dashed, $\theta_i=-34^\circ$) and best fit angles (black dashed, $\theta_i=-30.4^\circ$ and $\theta_r=30.4^\circ$).
\label{IntMach2244theta}
}
\end{figure}

The parameters ($\epsilon_i=0.171$ and $\theta_i=-30.4^\circ$) used in the model are different to that of the prescribed oblique soliton 
 {
($\epsilon_0=0.2$ and $\theta_0=-34^\circ$). 
}%
They are deduced from the stereoscopic reconstruction and appear to reflect the proper balance between the actual diffracting oblique solitons (incident and reflected) and the stem at the wall (see Fig.\,\ref{IntMach2244theta}).
With these values, $\kappa=1.03$ and the amplification factor $\epsilon_w/\epsilon_i=3.22$.

\begin{figure}
{{\includegraphics[width=127mm,viewport=95 370 550 490,clip]{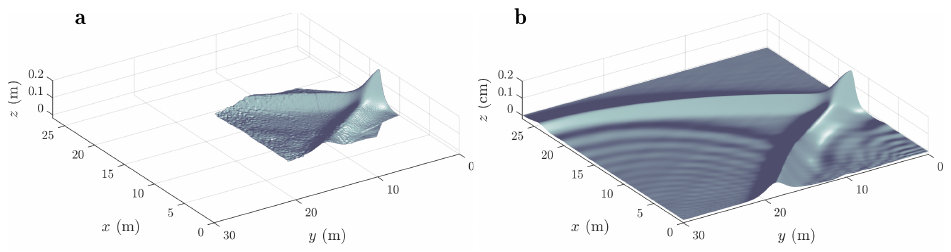}} }
\caption{
Snapshot of the free surface in the three-wave interaction {3S} measured by stereoscopy ({\bf a}) and corresponding simulation ({\bf b}).
\label{IntMachExpTolosa}
}
\end{figure}

This example shows that it is possible to produce a 2D soliton (a hump localized in both $x$ and $y$ directions) with very stable length and width. By adjusting the parameters of the experiment, it has been possible to compensate for diffraction effects in a zone covering the soliton's propagation over ten times its characteristic width. 
Such an observation gives credit to the second-order KP theory. 
The {Tolosa-lct} simulation reproduces it rather well, see Fig.\,\ref{IntMachExpTolosa}. The simulation also provides details of the dispersive waves produced during soliton generation.

\subsection{Simulations and asymptotic extrapolations  {in time}}

Only numerical simulation is capable of producing solitons in a domain large enough to avoid diffraction effects.
Moreover, wave dissipation can be controlled through the bottom friction parametrization. 
Simulations with {Tolosa-lct} in a $80\times 80$\,m$^2$ numerical tank, canceling bottom friction, were carried out to produce Mach interactions. 

\begin{figure}
\centering
{{\includegraphics[width=127mm,viewport=107 340 524 510,clip]{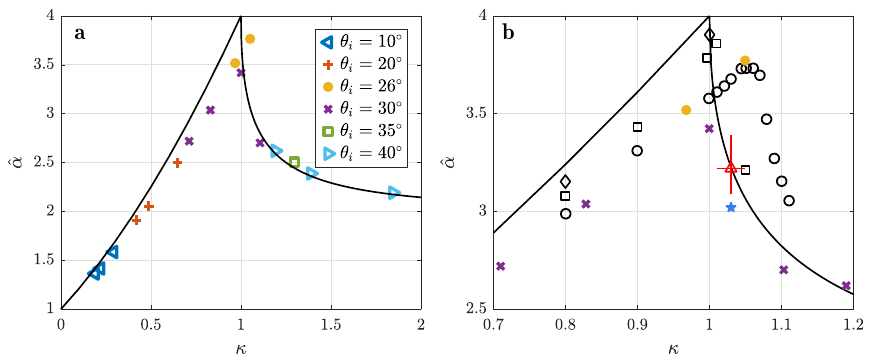}} }
\caption{
Stem amplification at the wall (\ref{kappaK}-\ref{alfachapo}) and comparison with \cite{kodama2016kp}'s theory (solid line). 
{\bf a} Extrapolated numerical results
with {Tolosa-lct} in a $80\times80$\,m$^2$ tank for various incidence angles and amplitudes 
 {and without bottom friction.}
{\bf b} 
Close-up around maximum resonance $\kappa=1$ and \cite{knowles2019fourfold}'s numerical results (blank symbols), for different angles: 
$\theta_i=10^\circ$ ({\scriptsize$\square$}),
$\theta_i=15^\circ$ ($\diamond$),
$\theta_i=26^\circ$ ($\circ$).
The quasi-steady 3-wave solution obtained with {Tolosa-lct} 
 {with bottom friction}
(blue star) and measured 3S (red triangle) are also plotted. For the latter, the error bars depict the variability with $\epsilon_i$ ($5.7\leq\epsilon_i\leq 6.25$\,cm from Fig.\,\ref{MachInt_rayleigh2244}{\bf c}).
\label{alfaTolosa}
}
\end{figure}

The results of these simulations show that the stem requires a long establishment distance when $\kappa\sim 1$, at the transition between irregular and regular interactions, in agreement with \cite{li2011mach} and \cite{knowles2019fourfold}. In the irregular case, the stem rapidly acquires an equilibrium velocity that is that of the soliton obtained asymptotically in the steady state \citep{leduquePhD2024}.
This makes it possible to extrapolate the results of the simulations and thus verify the results obtained by \cite{knowles2019fourfold} for amplification at the wall, 
see Fig.\,\ref{alfaTolosa}.
The agreement is rather good with the second-order KP theory proposed by \cite{kodama2016kp}, albeit indecisive close to resonance $\kappa\sim 1$, with notable discrepancies for small variations of the incidence angle. 
In this context, the stability conditions for the 3S solution shown in Figs.\,\ref{Tolosa2244}-\ref{IntMachExpTolosa} remain to be fully specified. 
 {
The measured amplification at the wall (red triangle in Fig.\,\ref{alfaTolosa}{\bf b}) and
 the simulation with bottom friction (blue star) are reasonably close to the inviscid second order KP theory (solid line).
Of note, the simulations without bottom friction at $\theta_i=30^\circ$ (crosses) also lie slightly below the curve. 
As a consequence, the bottom friction is likely not the dominant parameter that prevent to catch up the fourfold amplification at $\kappa = 1$.
Our experimental and computing constraints limited the exploration of 
$\kappa\sim 1$. We nevertheless confirm that near-resonant solutions are very sensitive to the main parameters ($\epsilon_i$, $\theta_i$) through the parameter $\kappa$ (\ref{kappaK}). 
}

\section{Conclusions} 

Experiments have been carried out in the large {\sc lhf} ({\sc Artelia}) wave tank to study the interactions of strongly non-linear waves. Our experiments 
are the first to detail soliton diffraction, studied theoretically recently by \cite{ryskamp2021evolution}. Stereoscopic reconstruction of free-surface displacements is achieved with good accuracy 
 {(of a few mm)}
over a large area of the tank ($\sim 180$\,m$^2$, or $\sim 320$\,m$^2$ using symmetrical experiments).

Diffraction plays a major role in our experiments. This is the cause of the main uncertainty in computing the $\kappa$ parameter, where $\kappa=1$ would be the maximum resonance with a fourfold amplification of the soliton amplitude in a Mach interaction at the tank wall. 

Validation of the {Tolosa-lct} code on a few cases enables us to extend the study to configurations that cannot be realized experimentally. This joint approach opens up the discussion of the relative stability of interacting solitons. 
In particular, in the vicinity of $\kappa=1$, we have shown on an example that a relatively stable 2D three-solitons solution can be generated, somehow confirming the predictions of \cite{kodama2016kp}, among others.

The properties of the solutions experimentally confirmed are also helpful to understand the so-called "solition-gas" state in 2D, in which a large number of solitons constantly interact generating stems that outlive the incident solitons. Predicted theoretically by \cite{zakharov1971kinetic}, such a state has only recently been obtained experimentally in 2D shallow water by \cite{leduque2024,leduque2025} and presented in detail by \cite{leduquePhD2024}. 

\subsection*{Acknowledgements}
This study was funded by the French National Research Agency (project {\sc sogood}, ANR-21-CE30-0061) and the Simons Foundation (Simons Collaboration on Wave Turbulence).
The Tolosa collection of software suite is developed by the {\it Institut de Math\'ematiques de Toulouse} as part of the national plan for wave-submersion vigilance, with funding from {\sc shom}. 
The technical support of {\sc artelia} and {\sc legi} was essential to the success of this project, in particular thanks to H. Michot and M. Lagauz\`ere on wave generation and F. Bonnel (Grenoble-INP) for capturing images.  
HM is deeply indebted to Lili Kimmoun who made some very significant contributions to a French project 
({\sc patom}) on wave breaking in the late 90's and 2000's. 


\begin{thebibliography}{}


\bibitem[Bonneton et~al., 2011]{bonneton2011splitting}
 {Bonneton, P., Chazel, F., Lannes, D., Marche, F., and Tissier, M. (2011).
\newblock A splitting approach for the fully nonlinear and weakly dispersive
  {G}reen--{N}aghdi model.
\newblock {\em J. Comput. Phys.}, 230(4):1479--1498.}

\bibitem[Chabchoub et~al., 2013]{chabchoub2013experimental}
Chabchoub, A., Kimmoun, O., Branger, H., Hoffmann, N., Proment, D., Onorato,
  M., and Akhmediev, N. (2013).
\newblock Experimental observation of dark solitons on the surface of water.
\newblock {\em Phys. Rev. Lett.}, 110(12):124101.

\bibitem[Choi et~al., 2022]{choi2022numerical}
Choi, J., Bokhove, O., Kalogirou, A., and Kelmanson, M. (2022).
\newblock Numerical experiments on extreme waves through oblique--soliton
  interactions.
\newblock {\em Water Waves}, 4(2):139--179.

\bibitem[Copie et~al., 2025]{copie2025controlled}
Copie, F., Suret, P., and Randoux, S. (2025).
\newblock Controlled manipulation of solitons in a recirculating fiber loop
  using external potentials.
\newblock {\em Phys. Rev. Res.}, 7(4):043232.

\bibitem[Fache et~al., 2024]{fache2024interaction}
Fache, L., Bonnefoy, F., Ducrozet, G., Copie, F., Novkoski, F., Ricard, G.,
  Roberti, G., Falcon, E., Suret, P., El, G., et~al. (2024).
\newblock Interaction of soliton gases in deep-water surface gravity waves.
\newblock {\em Phys. Rev. E}, 109(3):034207.

\bibitem[Funakoshi, 1980]{funakoshi1980reflection}
Funakoshi, M. (1980).
\newblock Reflection of obliquely incident solitary waves.
\newblock {\em J. Phys. Soc. J.}, 49(6):2371--2379.

\bibitem[Guizien and Barth\'elemy, 2002]{Guizien2}
Guizien, K. and Barth\'elemy, E. (2002).
\newblock Accuracy of solitary wave generation by a piston wave maker.
\newblock {\em J. Hydraul. Res.}, 40(3):321--331.

\bibitem[Kadomtsev and Petviashvili, 1970]{kadomtsev1970stability}
Kadomtsev, B.~B. and Petviashvili, V.~I. (1970).
\newblock On the stability of solitary waves in weakly dispersing media.
\newblock In {\em Doklady Akademii Nauk}, volume 192, pages 753--756. Russian
  Academy of Sciences.

\bibitem[Knowles and Yeh, 2019]{knowles2019fourfold}
Knowles, J. and Yeh, H. (2019).
\newblock Fourfold amplification of solitary-wave {M}ach reflection at a
  vertical wall.
\newblock {\em J. Fluid Mech.}, 861:517--523.

\bibitem[Kodama, 2010]{kodama2010kp}
Kodama, Y. (2010).
\newblock {KP} solitons in shallow water.
\newblock {\em J. Phys. A: Math. Theor.}, 43(43):434004.

\bibitem[Kodama and Yeh, 2016]{kodama2016kp}
Kodama, Y. and Yeh, H. (2016).
\newblock The {KP} theory and {Mach} reflection.
\newblock {\em J. Fluid Mech.}, 800:766--786.

\bibitem[Korteweg and deVries, 1895]{Korteweg1895}
Korteweg, D.~J. and deVries, G. (1895).
\newblock On the change of form of long waves advancing in a rectangular
  channel, and on a new type of long stationary waves.
\newblock {\em Phil. Mag.}, 39(5):422--443.

\bibitem[Leduque, 2024]{leduquePhD2024}
Leduque, T. (2024).
\newblock {\em \'Etude exp\'erimentale et num\'erique d'ondes non-lin\'eaires
  multidirectionnelles en eau peu profonde: application aux gaz de solitons}.
\newblock PhD thesis, Univ. Grenoble Alpes.

\bibitem[Leduque et~al., 2024]{leduque2024}
Leduque, T., Barth\'elemy, E., Michallet, H., Sommeria, J., and Mordant, N.
  (2024).
\newblock Space-time statistics of {2D} soliton gas in shallow water studied by
  stereoscopic surface mapping.
\newblock {\em Exp. Fluids}, 65(84).

\bibitem[Leduque et~al., 2025]{leduque2025}
Leduque, T., Kaczmarek, M., Michallet, H., Barth\'elemy, E., and Mordant, N.
  (2025).
\newblock From deep to shallow water {2D} wave turbulence: {E}mergence of
  soliton gas.
\newblock {\em Phys. Rev. Fluids}, 10(11):114801.

\bibitem[Li et~al., 2011]{li2011mach}
Li, W., Yeh, H., and Kodama, Y. (2011).
\newblock On the {M}ach reflection of a solitary wave: revisited.
\newblock {\em J. Fluid Mech.}, 672:326--357.

\bibitem[Miles, 1977a]{miles1977obliquely}
Miles, J.~W. (1977a).
\newblock Obliquely interacting solitary waves.
\newblock {\em J. Fluid Mech.}, 79(1):157--169.

\bibitem[Miles, 1977b]{miles1977resonantly}
Miles, J.~W. (1977b).
\newblock Resonantly interacting solitary waves.
\newblock {\em J. Fluid Mech.}, 79(1):171--179.

\bibitem[Nicolas, 1993]{nicolas1993}
Nicolas, P. (1993).
\newblock {\em Utilisation de variables distordues dans la th\'eorie des ondes
  de gravit\'e: application \`a la r\'efraction et \`a la g\'en\'eration}.
\newblock PhD thesis, Univ. Joseph Fourier, Grenoble.

\bibitem[Osborne, 2010]{Osborne}
Osborne, A.~R. (2010).
\newblock 30 - nonlinearity in duck pier data.
\newblock In Osborne, A.~R., editor, {\em Nonlinear Ocean Waves and the Inverse
  Scattering Transform}, volume~97 of {\em International Geophysics}, pages 779
  -- 794. Academic Press.

\bibitem[Perroud, 1957]{perroud1957solitary}
Perroud, P.~H. (1957).
\newblock {\em The solitary wave reflection along a straight vertical wall at
  oblique incidence}.
\newblock PhD thesis, Univ. California.

\bibitem[Rayleigh, 1876]{rayleigh1876}
Rayleigh, J. W.~S. (1876).
\newblock On waves.
\newblock {\em Philos. Mag}, 5(1):257--79.

\bibitem[Redor, 2019]{redor2019}
Redor, I. (2019).
\newblock {\em \'Etude exp\'erimentale de la turbulence int\'egrable en eau peu profonde}.
\newblock PhD thesis, Univ. Grenoble Alpes.

\bibitem[Redor et~al., 2020]{redor2020}
Redor, I., Barth\'elemy, E., Mordant, N., and Michallet, H.
  (2020).
\newblock Analysis of soliton gas with large-scale video-based wave measurements.
\newblock {\em Exp. Fluids}, 61(10).

\bibitem[Renouard et~al., 1985]{Renouard1985}
Renouard, D.P., Seabra-Santos, F.J., and Temperville, A.M. (1985).
\newblock {Experimental study of the generation, damping, and reflexion of a solitary wave}.
\newblock {\em Dyn. Atmos. Oceans}, 9:341--356.

\bibitem[Richard, 2021]{richard_Tolosa}
Richard, G.~L. (2021).
\newblock {An extension of the Boussinesq-type models to weakly compressible
  flows}.
\newblock {\em Eur. J. Mech.-B/Fluids}, 89:217--240.

\bibitem[Russell, 1845]{russell1845report}
Russell, J.~S. (1845).
\newblock {\em Report on Waves: Made to the Meetings of the British Association
  in 1842-43}.

\bibitem[Ryskamp et~al., 2021]{ryskamp2021evolution}
Ryskamp, S., Maiden, M.~D., Biondini, G., and Hoefer, M.~A. (2021).
\newblock Evolution of truncated and bent gravity wave solitons: the {Mach}
  expansion problem.
\newblock {\em J. Fluid Mech.}, 909:A24.

\bibitem[Tanaka, 1993]{tanaka1993mach}
Tanaka, M. (1993).
\newblock Mach reflection of a large-amplitude solitary wave.
\newblock {\em J. Fluid Mech.}, 248:637--661.

\bibitem[Wang and Pawlowicz, 2012]{wang2012oblique}
Wang, C. and Pawlowicz, R. (2012).
\newblock Oblique wave-wave interactions of nonlinear near-surface internal
  waves in the strait of georgia.
\newblock {\em Journal of Geophysical Research: Oceans}, 117(C6).

\bibitem[Zabusky and Kruskal, 1965]{zabusky1965interaction}
Zabusky, N.~J. and Kruskal, M.~D. (1965).
\newblock Interaction of ``solitons" in a collisionless plasma and the
  recurrence of initial states.
\newblock {\em Phys. Rev. Lett.}, 15(6):240.

\bibitem[Zakharov, 1971]{zakharov1971kinetic}
Zakharov, V. (1971).
\newblock Kinetic equation for solitons.
\newblock {\em Sov. Phys. JETP}, 33(3):538--540.

\end{thebibliography}

\end{document}